\documentclass[aps,prb,twocolumn,showpacs,amsmath,amssymb]{revtex4-1}
\usepackage{graphicx}
\usepackage{dcolumn}
\usepackage{epstopdf}
\usepackage{bm}
\usepackage{braket}
\usepackage[usenames,dvipsnames]{xcolor}
\usepackage[normalem]{ulem}
\usepackage{soul}

\begin{document}
\title{A high-temperature expansion method for calculating paramagnetic exchange interactions}
\author{O.M. Sotnikov$^{1}$, V.V. Mazurenko$^{1}$}
\affiliation{$^{1}$Theoretical Physics and Applied Mathematics Department, Ural Federal University, Mira Str.19,  620002 Ekaterinburg, Russia}
\date{\today}

\begin{abstract}
The method for calculating the isotropic exchange interactions in the paramagnetic phase is proposed.  It is based on the mapping of the high-temperature expansion of the spin-spin correlation function calculated for the Heisenberg model onto Hubbard Hamiltonain one. The resulting expression for the exchange interaction has a compact and transparent formulation. The quality  of the calculated exchange interactions is estimated by comparing the eigenvalue spectra of the Heisenberg model and low-energy magnetic part of the Hubbard model. By the example of quantum rings with different hopping setups we analyze the contributions from the different part of the Hubbard model spectrum to the resulting exchange interaction.

\end{abstract}
\maketitle

\section{Introduction}

The magnetic properties of a correlated system can be fully described by its magnetic susceptibility characterizing the response of the system to an external magnetic field.\cite{White}
Modern numerical methods of the dynamical mean-field theory for solving realistic electronic models provide the most reliable information concerning the electronic and magnetic excitation spectra of the strongly correlated materials. Importantly, by using the dynamical mean-field theory\cite{DMFT} (DMFT) the frequency- and momentum-dependent susceptibilities of a correlated material can be directly calculated at different external parameters (temperatures and magnetic fields) and compared with those measured in the experiment.
However, the solution and reproduction of the experimentally observed susceptibilities do not mean a truly microscopic understanding of the magnetic properties formation. 
In this respect the determination of the individual magnetic interactions, $J_{ij}$ of the Heisenberg model is of crucial importance. The corresponding Hamiltonian is given by
\begin{eqnarray}
\hat H_{Heis}= \sum_{ij} J_{ij} \hat {\vec S}_{i} \hat {\vec S}_{j}.
\end{eqnarray}

The development of the methods for calculating the exchange interactions $J_{ij}$ between magnetic moments  in modern materials is an active research field. \cite{Katsnelson, Fransson, Secchi, Liechtenstein} Some important examples of methods are listed in Table I.  The density-functional exchange formula proposed in Ref.\onlinecite{Liechtenstein} is based on the idea about infinitesimal rotation of the magnetic moments from the collinear ground state. The resulting exchange interaction is the response of the system on this perturbation. Being formulated in terms of the Green's function of the system such an approach has a number of important options, for instance, it is possible to calculate the orbital contributions to the total exchange interaction. The latter opened a way for a truly microscopic analysis of the magnetic couplings. 

Then in Ref.\onlinecite{Katsnelson} the method for calculating magnetic couplings within the LDA+DMFT scheme was reported. Such an approach facilitates the analysis of the exchange interactions taking the dynamical Coulomb correlations into account.\cite{Kvashnin1, Kvashnin2, Eriksson, SolovyevCrO2, Wan}  Recently, a general technique to extract the complete set of the magnetic couplings by taking into account the vertices of two-particle Green's functions and non-local self-energies was developed in Ref.\onlinecite{Secchi}. 

By construction the methods reviewed above assume some type of the magnetic ordering in the system. However, there are examples when the resulting exchange interactions are very sensitive to the particular magnetic configuration.\cite{SolovyevCrO2} Thus one may obtain different sets of the magnetic couplings for the same system. 

\begin{table}
\caption[Bset]{\label{usualMethods}List of methods for calculation of the isotropic exchange interaction. $\phi_{i}(x)$ is a wave function centered at the lattice site i. $t_{ij}$ and $U$ are the hopping integral and the on-site Coulomb interaction, respectively. $z$ is the number of the nearest neighbors. $E_{FM}$ and $E_{AFM}$ are the energies of the ferromagnetic and antiferromagnetic solutions obtained by using a mean-field electronic structure approach. $\Gamma^{(1)}_{ij}$ is the first order term in the high-temperature expansion of the spin-spin correlation function.}
\begin{tabular}{cc}
\hline
Method & Expression \\
\hline\\
Heitler-London's exchange\cite{Heis} & $J_{ij} = \int\frac{\phi^*_i(x)\phi_j(x)\phi^*_j(x')\phi_i(x')}{|x-x'|}dxdx'$ \\
\\
Anderson's superexchange & $J^{kin}_{ij} = \frac{4t^2_{ij}}{U}$ \\
theory\cite{Anderson} & \\
\\
Total energies method & $J = \frac{E_{FM}-E_{AFM}}{4zS^2}$ \\
\\
Local force theorem\cite{Liechtenstein} & $J_{ij} = \frac{\partial^2 E}{\partial \vec S_{i} \partial \vec S_{j}}$ \\
\\
HTE method (this work) & $J_{ij} = - \frac{\Gamma^{(1)}_{ij}}{(\frac{1}{3}S(S+1))^2}$ \\ 
\\
\hline
\end{tabular}
\end{table}

Another important methodological problem in this research field concerns the determination of the magnetic couplings in a system being in a disordered magnetic phase. Numerous magnetic experiments \cite{SrCuBO3,Tsirlin11,Tsirlin12} revealed quantum spin systems which, due to the low-dimensional crystal structure, do not exhibit any sign of the magnetic ordering even at {\it very low temperatures}. Since the electron hopping integral, $t_{ij}$ in these materials is much smaller than the on-site Coulomb interaction, $U$, then  the magnetic coupling can be associated with the Anderson's superexchange interaction,\cite{Anderson} $J_{ij}=\frac{4t^2_{ij}}{U}$. For intermediate values of $t/U$ ($\sim$ 0.1) one can still use the pure spin model with parameters defined from the high-order strong coupling expansion within perturbative continuous unitary transformations.\cite{Stein,Mila}

In turn, the simulation of the exchange interactions in high-temperature paramagnetic phases can be performed by means of the dynamical mean-field theory and its extension. For instance, in case of the $\gamma$-iron  the authors of Ref.\onlinecite{Katanin} compared the magnetic susceptibilities obtained for Heisenberg model within 1/$z$ expansion and that calculated in DMFT approach. To describe the formation of the local magnetic moment and exchange interaction in the $\alpha$-iron, a spin-fermion model was proposed in Ref.\onlinecite{Katanin1}.  

Here we report on a distinct method, high-temperature expansion (HTE) method for calculating the isotropic exchange interactions in the paramagnetic phase. It is based on the mutual mapping of the high-temperature spin-spin correlation functions calculated in Hubbard and Heisenberg models.  Being formulated for finite clusters our method can be applied to the investigation of the magnetic couplings in magnetic molecules or nanostructures deposited on the insulating and metallic surfaces. It can be also expanded on the calculation of the high-order couplings such as ring exchange. We have used the developed approach to study the magnetic interactions in quantum spin rings with different hopping configurations.    

\section{Methods}

The main focus in our approach is concentrated on the spin-spin correlation function,
\begin{eqnarray}
 \Gamma_{ij} = \frac{{\rm Tr}(\hat{S}^z_i\hat{S}^z_je^{-\beta\hat{H}})}{{\rm Tr}(e^{-\beta\hat{H}})},
\end{eqnarray}
where $\beta$ is the inverse temperature and $\hat H$ is the Hamiltonian describing the system in question. Since the paramagnetic regime is of our interest, then we can consider the $z$-component of the spin operator.
Following Ref.\onlinecite{Ashcroft} we consider $\Gamma_{ij}$ in the high temperature limit in which the exponent is expanded as $e^{-\beta\hat{H}}= 1- \beta \hat H$. Thus one obtains
 \begin{eqnarray}
 \Gamma_{ij} \approx \Gamma_{ij}^{(0)} + \beta\Gamma_{ij}^{(1)} = \frac{{\rm Tr}(\hat{S}^z_i\hat{S}^z_j) - \beta {\rm Tr} (\hat{S}^z_i\hat{S}^z_j \hat H)}{{\rm Tr}(1) - \beta {\rm Tr}(\hat H)},
 \label{Gamma}
\end{eqnarray}
where ${\rm Tr} (\hat A)$ is the trace that corresponds to the summation over all eigenstates of the Hamiltonian of the system, $\hat H$,
\begin{eqnarray}
\label{Trace}
{\rm Tr (\hat A)} = \sum_{n} \langle \Psi_{n} |\hat A| \Psi_{n}\rangle.
\end{eqnarray}

If the system in question can be described by the Heisenberg Hamiltonian with localized magnetic moments then in zero order on $\beta$ one obtains
\begin{eqnarray}
 \label{eq:spszsz}
 {\rm Tr}(\hat{S}_i^z\hat{S}_j^z) = \frac{1}{3}S(S+1)N\delta_{ij},
\end{eqnarray}
which simply means that the spins are independent at high temperatures. Here \mbox{$N = (2S+1)^L$} is the number of the eigenstates of the Heisenberg Hamiltonian ($L$ denotes the number of sites in the model).

The same idea is used when analyzing the contribution of the first order on $\beta$ to the spin-spin correlation function that carries the information concerning the exchange interaction between the spins,
\begin{eqnarray}
 \label{eq:spszszH}
 {\rm Tr}(\hat{S}_i^z\hat{S}_j^z\hat{H}_{Heis}) = {\rm Tr}(\hat{S}_i^z\hat{S}_j^z\sum_{m \neq n}J_{mn}\hat{\vec{S}}_m\hat{\vec{S}}_n) = \nonumber \\
 = J_{ij}{\rm Tr}(\hat{S}_i^z\hat{S}_j^z\hat{S}_i^z\hat{S}_j^z) = J_{ij}N\left(\frac{1}{3}S(S+1)\right)^2. \label{TrHeis}
\end{eqnarray}

This high-temperature decomposition of the spin-spin correlation function was used by the authors of Ref.\onlinecite{Ashcroft} to obtain the expression for the Curie-Weiss temperature. As we will show below it can be also used for calculating $J_{ij}$. 

In the seminal work by Anderson \cite{Anderson} the Heisenberg exchange interaction is defined in terms of the Hubbard model parameters, $t_{ij}$ and $U$. For that the author considered the limit $t_{ij} \ll U$, in which one can obtain the famous superexchange expression, $J_{ij} = \frac{4t_{ij}^2}{U}$.  

Our method for calculating $J_{ij}$ is also based on the using of the Hubbard model that in the simplest one-band form can be written as
\begin{eqnarray}
\hat H_{Hubb}=\sum_{ij \sigma} t_{ij} \hat c^{+}_{i \sigma} \hat c_{j \sigma} + \frac{U}{2} \sum_{i \sigma} \hat n_{i \sigma} \hat n_{i -\sigma} - \mu \sum_{i \sigma} \hat n_{i \sigma},
\label{Hubbard}
\end{eqnarray} 
where $\sigma$ is the spin index, $t_{ij}$ is the hopping integral between i$th$ and j$th$ sites, $U$ is the on-site Coulomb interaction and $\mu$ is the chemical potential.

Since our aim is to define the parameters of the Heisenberg model with localized spins, on the level of the Hubbard model it is naturally to start with the atomic limit in which the hopping integral is much smaller than the Coulomb interaction, $U \gg t$. In this case the spectrum of the eigenvalues can be divided onto low- and high-energy parts that are related to the 
magnetic excitations of the Heisenberg type and charge excitations of the order of $U$, respectively.
Our method is based on the comparison of the magnetic observables such as spin-spin correlation functions calculated in Hubbard model and Heisenberg model approaches in the high temperature limit, $\beta \rightarrow$ 0. 

In general, the trace over spin operators, Eq.(\ref{Trace}) differs in the case of the Heisenberg and Hubbard models. For instance, one should perform the summation over all eigenstates for the Heisenberg model. At the same time in the case of the Hubbard model one should exclude the high-energy eigenstates with doubly occupied sites from the consideration. The energies of these states are of order of $U$. 

In the limit of the localized spins $t \ll U$ that we consider, the traces ${\rm Tr} (\hat S^z_{i} \hat S^z_{j})$ (in the numerator of Eq.(\ref{Gamma})) and ${\rm Tr} (\hat H)$ (in the denominator of Eq.(\ref{Gamma})) are similar to that defined for the Heisenberg model. 

We are interested in the first order term on the inverse temperature for which one obtains
\begin{eqnarray}
{\rm Tr} (\hat S^z_{i} \hat S^z_{j} \hat H_{Hubb}) = \sum_{n=0}^{N-1} \langle \Psi_{n} | \hat S^z_{i} \hat S^z_{j} \hat H_{Hubb} | \Psi_{n} \rangle = \nonumber \\
= \sum_{n=0}^{N-1} E_{n} \langle \Psi_{n} | \hat S^z_{i} \hat S^z_{j}| \Psi_{n} \rangle{}, \label{TrHubb}
\end{eqnarray}
here $E_{n}$ is the eigenvalue of the Hubbard model, $\Psi_{n}$ is the corresponding eigenvector and $N$ is the number of the eigenstates of the Heisenberg Hamiltonian.

Comparing Eq.(\ref{TrHeis}) and Eq.(\ref{TrHubb}) one can derive the following expression for the Heisenberg's exchange interaction
\begin{eqnarray}
 \label{eq:J}
 J_{ij} = \frac{\sum_{n=0}^{N-1} E_n \braket{\Psi_n | \hat{S}^z_i\hat{S}^z_j | \Psi_n}} {N (\frac{1}{3}S(S+1))^2}.
\end{eqnarray}

Let us analyze the obtained expression for the paramagnetic exchange interaction. First of all, it contains the summation over all eigenstates belonging to the magnetic part of the full Hubbard spectrum. The high-energy part of the Hubbard spectrum describing the charge excitations is excluded from the consideration. For each eigenstate we measure the correlation between two spins. Such a correlation can be positive or negative depending on the spin configuration encoded in the eigenstate and is multiplied by the excitation energy with respect to the ground state with $E_{0}=0$.

The expression for the exchange interaction Eq.(\ref{eq:J}) was obtained by comparing the spin-spin correlation functions of the Heisenberg and Hubbard models in the limit $t \ll U$. Despite of this, in some cases the calculated exchange interactions, as we will show below, lead to good agreement of the Heisenberg and Hubbard eigenvalue spectra even for $\frac{t}{U} \sim 1$. Importantly, one can analyze the magnetic interactions in the strongly correlated regime. 

For transition metal oxides the typical ratio between hopping integral and on-site Coulomb interaction is of order of 0.03. It was shown that in case of $5d$ iridium oxides \cite{SolovyevBa214} this value can be about two times larger, 0.07 and the implementation of the ordinary superexchange theory is questionable. 
The simulation of the magnetic interaction in metallic systems is another complicated problem, we deal with the situation when the hopping integrals are of the same order of magnitude as the Coulomb interaction.  

In the case of the many-band Hubbard model the spin operator of the i$th$ site in Eq.(\ref{eq:J}) can be written as the sum of the orbital contributions, $\hat S^{z}_{i} = \sum_{m} \hat S^{z}_{i,m}$. Thus for the $S>\frac{1}{2}$ we obtain the following expression for the paramagnetic exchange interaction
\begin{eqnarray}
 J_{ij} = \frac{\sum_{mm'} \sum_{n=0}^{N-1} E_n \braket{\Psi_n | \hat{S}^z_{i, m} \hat{S}^z_{j, m'} | \Psi_n}}{N(\frac{1}{3}S(S+1))^2}.
\end{eqnarray}

One important problem when calculating the exchange interaction is how to estimate and control the quality of the obtained exchange interactions. It can be done by solving the corresponding Heisenberg model and by calculating the experimentally observed quantities (such as magnetic susceptibility, magnetization and other). The comparison of the calculated theoretical dependencies with the available experimental data is standard way to define the reability of the  constructed Heisenberg model. In our study the exchange interactions for the spin model estimated on the basis of the electronic Hubbard Hamiltonain. Thus it is natural to estimate the quality of the constructed Heisenberg model by comparing the eigenvalue spectra of the spin model and parent electronic Hamiltonian at different degree of the localization.  

\section{Exact solution for dimer}
\label{dimerSec}
The electronic and magnetic excitation spectra of the dimer that can be obtained analytically is the classical test in the field of the strongly correlated materials. Importantly, there are a lot of examples of the real low-dimensional materials that have the dimer motif.\cite{dimer1,dimer2,dimer3} The superexchange interaction in the dimer can be also simulated within the experiments with ultracold atoms in optical lattice.\cite{real-time-exch}. In such experiments the hopping integral and on-site Coulomb interaction can be varied in a wide range. For instance, the authors of Ref.\onlinecite{Greif} explored the ratios ranging from the metallic ($t/U \sim 0.1$) to insulating ($t/U \ll 1$) regimes  when performed the quantum simulations on the two-dimensional optical lattice.

Within the proposed method we are interested in \mbox{$N=4$} lowest magnetic eigenstates of the Hubbard model, they are presented in Table~\ref{dimerHubb}, where the following notations are used \mbox{$C_0^2 = \frac{1}{2(1+\epsilon_-^2)}$}, \mbox{$\epsilon_- = \frac{U(1-\gamma{})}{4t}$}, \mbox{$\gamma = \sqrt{1+\frac{16t^2}{U^2}}$}. The eigenvalues are the following: \mbox{$E_0 = \frac{U}{2}(1-\gamma)-2\mu$}, \mbox{$E_1 = E_2 = E_3 = -2\mu$}.

\begin{table}[!h]
\caption[Bset]{Four lowest eigenstates of the Hubbard model for the dimer.}
\label {dimerHubb}
\begin {tabular}{c|c}
\hline
$n$ & $\Psi_n$ \\
\hline\\[-2mm]
0 & [($|$ \st{\,$\downarrow$\,} \st{\,$\uparrow$\,} $\rangle - |$ \st{\,$\uparrow$\,} \st{\,$\downarrow$\,} $\rangle) + \frac{U(1-\gamma{})}{4t}(|$ \st{$\downarrow\uparrow$} \st{~\,~} $\rangle + |$ \st{~\,~} \st{$\downarrow\uparrow$} $\rangle)]\cdot{}C_0$ \\ [2mm]
1 & $|$ \st{\,$\uparrow$\,} \st{\,$\uparrow$\,} $\rangle$ \\ [2mm]
2 & $|$ \st{\,$\downarrow$\,} \st{\,$\downarrow$\,} $\rangle$ \\ [2mm]
3 & $\frac{1}{\sqrt{2}}\cdot{}(|$ \st{\,$\uparrow$\,} \st{\,$\downarrow$\,} $\rangle$ + $|$ \st{\,$\downarrow$\,} \st{\,$\uparrow$\,} $\rangle$) \\ [2mm]
\hline
\end {tabular}
\end {table}

By using the developed method Eq.\eqref{eq:J}, we obtain the following exchange interaction in the dimer
\begin{eqnarray}
 \label{eq:dimerJ}
 J = -\frac{U}{2}\left(1-\gamma\right). 
\end{eqnarray}
This value is exactly the excitation energy from the singlet to triplet state of the dimer, $E_1 - E_0$. Thus our method can be used to construct a Heisenberg model reproducing the Hubbard model spectrum for any reasonable ratio of kinetic and Coulomb interaction parameters, $\frac{t_{ij}}{U}$. On the other hand the Heisenberg model constructed by means of the Anderson's superexchange theory, $J_{ij} = \frac{4t_{ij}^{2}}{U}$ results in the spectrum deviating from that of the original Hubbard model at $\frac{t_{ij}}{U} >$ 0.2.

\begin{figure}
\includegraphics[angle=0,scale=1]{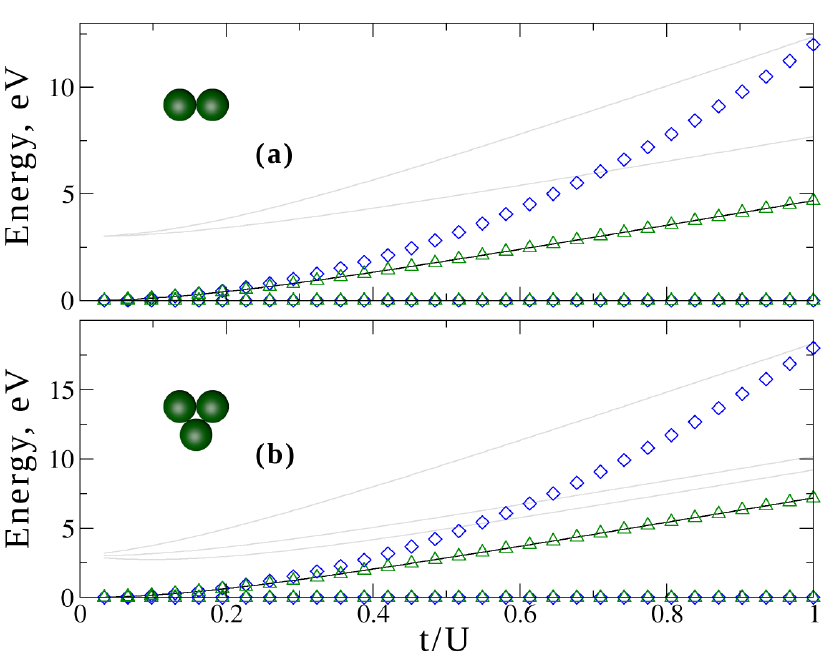}
\caption {Comparison of the excitation spectra of the Hubbard model (solid lines) and Heisenberg models with exchange interaction calculated by using the developed method (green triangles) and superexchange Anderson approach (blue rhombus)}\label{dimertriangle}
\end{figure}

\subsection{Comparison with the Hartree-Fock solution}
One of the important results of modern magnetism theory was the development of the local force theorem\cite{Liechtenstein} for calculating the exchange interactions. Such an approach give reliable results and is widely used for simulation magnetic properties of the transition metal compounds.\cite{Kvashnin1, Kvashnin2, Eriksson, SolovyevBa214, SolovyevCrO2} Thus the next step of our investigation was to
compare the results of the high-temperature expansion method we developed and those obtained by using the density-functional exchange formula. For these purposes we have chosen the dimer system. 

Since the method based on the local force theorem requires a non-zero magnetization of the system we used the Hartree-Fock approximation to solve the Hubbard model, Eq.(\ref{Hubbard})
\begin{eqnarray} 
\hat{H}_{HF} = \sum\limits_{ij\sigma}t_{ij}\hat{c}^{+}_{i\sigma}\hat{c}_{j\sigma}
+U\sum_{i\sigma}\langle{}\hat{n}_{i-\sigma}\rangle{}\hat{n}_{i\sigma}.
\end{eqnarray}
According to the local force theorem the exchange interaction is given by
\begin{eqnarray}
\label{eq:Lichtenstein}
 J_{ij} = \frac{1}{2\pi{}S^2}\int\limits_{-\infty}^{E_F}{\rm Im}(\tilde V_i G_{ij}^\downarrow \tilde V_j G_{ji}^\uparrow)\ d\varepsilon,
\end{eqnarray}
here $E_{F}$ is the Fermi level, $\tilde V_i = V_{i}^\uparrow - V_{i}^\downarrow$ denotes the spin-dependent Hartree-Fock potential calculated self-consistently and $G^{\uparrow,\downarrow}(\varepsilon) = (\varepsilon - H_{HF}^{\uparrow,\downarrow})^{-1}$ is the Green's function of the Hartree-Fock Hamiltonian. Unlike our approach Eq.\eqref{eq:J}, this formula requires presence of a magnetic order in the system.

\begin{figure}
\includegraphics[angle=0,scale=1]{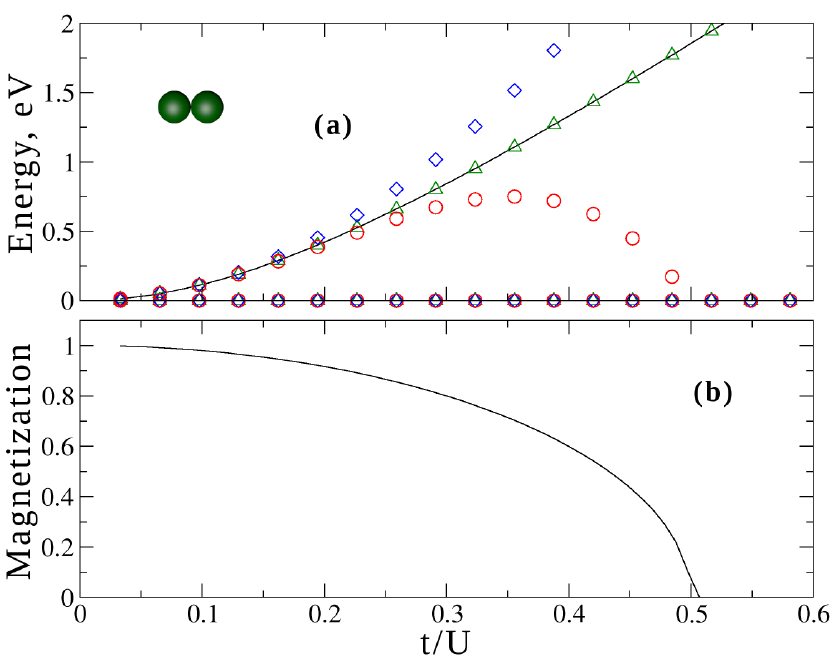}
\caption {(a) Comparison of the eigenvalue spectra obtained from the solution the Hubbard model (black solid line) and the Heisenberg models with parameters calculated by the developed approach Eq.(\ref{eq:J}) (green triangles), local force theorem method Eq.(\ref{eq:Lichtenstein}) (red circles) and Anderson's superexchange theory (blue rhombus). (b) Magnetization as a function of the localization.}\label{fig:hfspec}
\end{figure}

In the case of the dimer, the exchange interactions obtained by using local force approach give excellent agreement with the spectrum of the Hubbard model for $\frac{t}{U} < $0.2 (Fig.~\ref{fig:hfspec}). For larger values of the hopping integrals the averaged magnetic moment is strongly suppressed and becomes almost zero at $t/U \sim 0.5$. These results indicate the limits of the applicability of the mean-field Green's function approach for calculating the exchange interaction in strongly correlated systems.  

\section{Solutions for triangle and trimer}
{\it Triangle} is another example of the model for which we obtain excellent agreement of the electronic and spin eigenvalue spectra. The Heisenberg model spectrum for the triangle consists of  
four-fold degenerate ground and four-fold degenerate excited states. As in the case of the dimer, the exchange interaction between spins in the triangle is defined by the corresponding splitting between excited and ground state levels. From Fig.\ref{dimertriangle} the Heisenberg model, which we constructed by using the HTE method, precisely reproduces the magnetic part of the Hubbard model.   

\begin{figure}
\includegraphics[angle=0,scale=1.0]{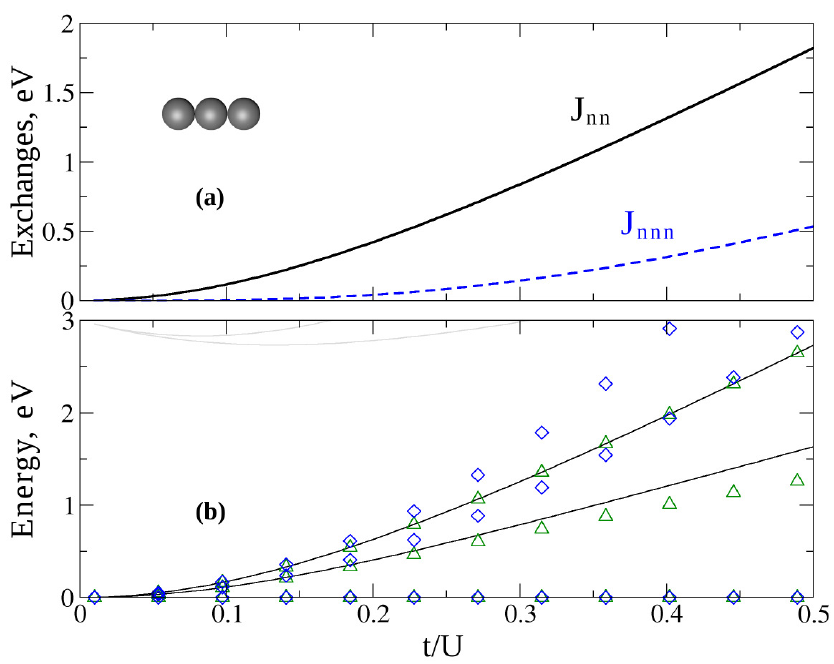}
\caption {(a) The calculated exchange interactions between nearest neighbours, J$_{nn}$ and next-nearest neighbours, J$_{nnn}$ in the trimer.  
(b) Comparison of the eigenvalue spectra for the trimer. Solid lines denote the Hubbard model spectrum. Blue rhombus, green triangles and red dashed lines correspond to The Heisenberg model solutions with different sets of the exchange interactions are presented by blue rhombus (the Anderson's superexchange) and green triangles (the developed HTE method, Eq.(\ref{eq:J})).}\label{fig:ideal}
\end{figure}

{\it Trimer.} The situation becomes more complicated if we consider the trimer with the nearest neighbor hopping presented in Fig.\ref{fig:ideal}. In contrast to the triangle the ground state of the trimer is two-fold degenerate. In turn the highest excited state in the magnetic part of the eigenspectrum is four-fold degenerate. As we will show below the two-fold intermediate excited level is related to the interaction between next nearest neighbours.  

For such hopping setup, within Anderson's superexchange theory we obtain the antiferromagnetic coupling, $J_{ij} = \frac{4t^2_{ij}}{U}$ between nearest neighbors in the trimer. To define the interaction between next nearest neighbors one should use the fourth-order perturbation theory on the hopping. The situation becomes more complicated if the condition $t_{ij} \ll U$ is not fulfiled.  On the other hand  by using the developed method Eq.(\ref{eq:J}) we obtain antiferromagnetic nearest and second nearest neighbors exchanges. The solution of the corresponding Heisenberg model leads to perfect agreement between the spin and Hubbard model spectra up to large values of the ratio $\frac{t_{ij}}{U}$.  

In the case of the trimer we can also explicitly relate the exchange interactions with the eigenvalues spectrum of the Hubbard model. For that we used the condition $E_{n}^{Hubb} - E_{n'}^{Hubb} = E_{n}^{Heis} - E_{n'}^{Heis}$ and obtained the following expressions for the magnetic couplings in the trimer:  
\begin{eqnarray}
 \label{eq:exact}
 \begin{split}
  &J_{nn} = \frac{2}{3}(E_4 - E_0) \\
  &J_{nnn} = J_{nn} - (E_2 - E_0)
 \end{split}
\end{eqnarray}
where $J_{nn}$ and $J_{nnn}$ are exchange interactions between nearest neighbors and next-nearest neighbors in the trimer, respectively. One can see that the leading exchange interaction $J_{nn}$ between the nearest neighbours is related to the energy splitting between ground state and highest excited state belonging to the magnetic part of the whole electronic spectrum. The situation with the next nearest-neighbour coupling is more complicated. In addition to the $E_{4}-E_{0}$ that is related to the leading exchange interaction, it also has the ferromagnetic contribution from the intermediate excited state, $E_{2}-E_{0}$. As we will show below the similar picture is observed in quantum spin rings. 

\section{Quantum spin rings}
In this section we present the results of computer simulations concerning the magnetic interactions in the finite quantum clusters with ring geometry. The theoretical interest in these systems is due to the synthesis and study of the magnetic properties of the molecular magnets with ring geometry.\cite{Fe-ring, Fe3, Schnack1, Schnack2} Such systems demonstrate a number interesting and complex phenomena, quantum spin tunneling, long-time spin relaxation, topological spin phases (Berry phases) and others. In this respect the microscopic understanding of the intra-molecular magnetic couplings plays a crucial role.\cite{MazurenkoMn12} On the other hand the spin rings are also of great practical interest, since they can be used as building elements for novel quantum communication technologies \cite{Bose} and for engineering quantum memory that is stable against noise and imperfections.\cite{Laser} 

\begin{figure}
\includegraphics[angle=0,scale=1]{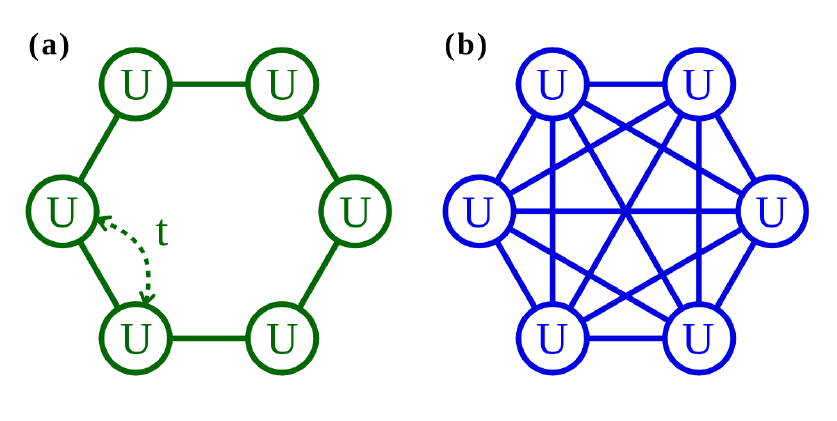}
\caption {Schematic representation of the hopping setups for the Hubbard Hamiltonian  simulations. (Left) The ring model with the nearest neighbor hoppings. (Right) All-to-all configuration in which all the hoppings between sites are the same.}\label{fig:rall-toall}
\end{figure}

In our study we have simulated the magnetic interactions in the quantum rings describing by the Hubbard models with different hopping setups presented in Fig.\ref{fig:rall-toall}.
They can be realized in the quantum simulation experiments on ultracold atoms in optical lattices.\cite{Greif1}

\subsection{Rings with nearest neighbors hoppings}

\begin{figure}
\includegraphics[angle=0,width=0.5\textwidth]{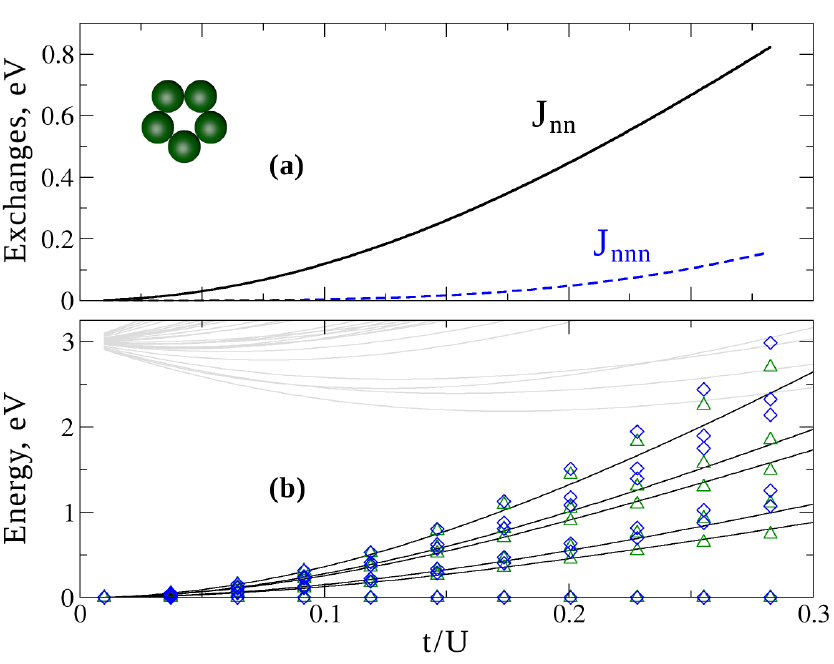}
\caption {(a)  The calculated exchange interactions between nearest neighbours, J$_{nn}$ and next-nearest neighbours, J$_{nnn}$ in the 5-site ring. (b) Comparison of the eigenvalue spectra for the 5-site ring. Solid lines denote the Hubbard model spectrum. Blue rhombus and green triangles correspond to the Heisenberg model solutions with different sets of the exchange interactions: blue rhombus (Anderson's superexchange) and green triangles (developed method, Eq.(\ref{eq:J})).}\label{my1}
\end{figure}

First, we analyze the results of the simulations for quantum rings describing the Hubbard model with the only nearest neighbor hopping integral.
Similar to the case of the dimer and trimer our method leads to better agreement between Heisenberg and low-energy Hubbard model spectra than the others.
Fig.\ref{my1} gives the comparison of the eigenvalues spectra calculated by different methods  in the case of the 5-site ring.
One can see that the high-temperature expansion method reproduces the electronic Hamiltonian spectrum up to $t/U$ = 0.28. At this value the high- and low-energy parts of the spectrum overlap, which prevents us from determining the exchange interaction.

In the hopping setup we used (Fig.\ref{fig:rall-toall}, left) there are hopping integrals between nearest neighbors only. Nevertheless each site has non-zero antiferromagnetic exchange interaction with all the other sites in the ring (we denote them J$_{nnn}$). Fig.\ref{my1}(a) demonstrates the behavior of such diagonal couplings at different $t/U$ ratios in comparison with the leading exchange interaction between nearest neighbors, J$_{nn}$. Despite of the fact that the coupling J$_{nnn}$ growths much slower than the nearest-neighbour one it cannot be neglected when constructing the Heisenberg model at $t/U >$ 0.15. It can be clearly seen from Fig.\ref{my1} (b), in which the Anderson's superexchange theory with zero J$_{nnn}$
leads to the eigenvalue spectrum deviating from the Hubbard model one.

\begin{figure}
\includegraphics[angle=0,scale=1]{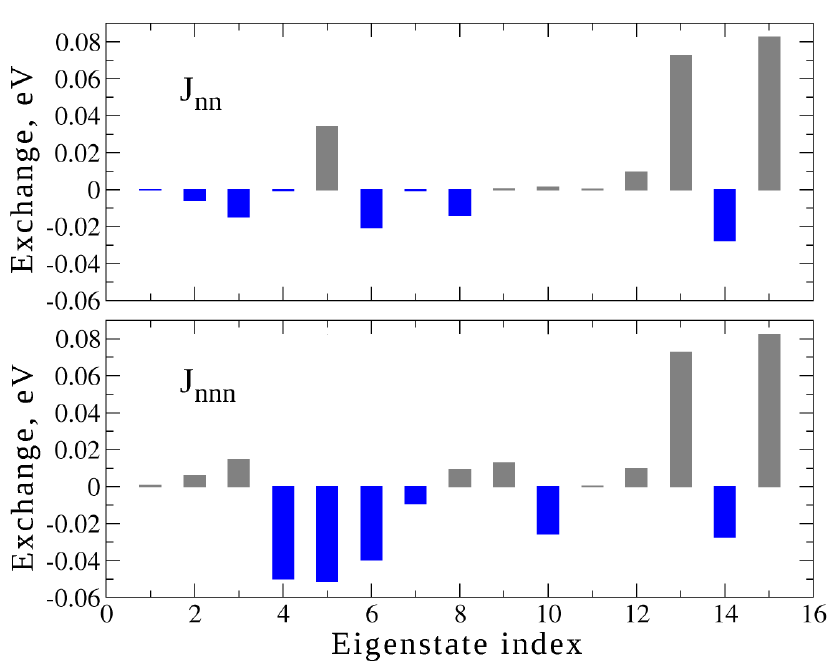}
\caption {Contributions from the different eigenstates of the Hubbard model to the nearest neighbor (top) and next-nearest neighbour (bottom) exchange interactions calculated for 4-site ring with $t/U$ = 0.1.}\label{my2}
\end{figure}

The expression for the paramagnetic exchange interaction, Eq.(\ref{eq:J}) that we derived contains the summation of the eigenstates belonging to the low-energy magnetic part of the Hubbard model spectrum. It is important to analyze the contribution of the individual eigenstates to the resulting exchange interaction. From Fig.\ref{my2} one can see that there are ferromagnetic contributions that partially compensate the antiferromagnetic ones. Interestingly, the contributions from the highest excited states are almost the same for the J$_{nn}$ and J$_{nnn}$ couplings.  As it was shown in the case of the trimer, the intermediate excited eigenstates produce the ferromagnetic contributions to J$_{nnn}$. 

\subsection{Rings with all-to-all hoppings.}
By the example of the results for the 5-site ring presented in Fig.\ref{my1} (b) one can see that the quantum rings with nearest neighbor hopping demonstrate rather complicated spectra. However, for practical purposes, for instance, to construct a quantum logic device, we need a system with the excitation spectrum as simple as possible.  
In the case of the quantum rings that we consider the excitation spectrum can be considerably simplified by introducing the same hopping integral for all the bonds in the quantum Hamiltonian. It is so-called all-to-all hopping configuration (Fig.\ref{fig:rall-toall}). 

\begin{figure*}
\includegraphics[angle=0,scale=1]{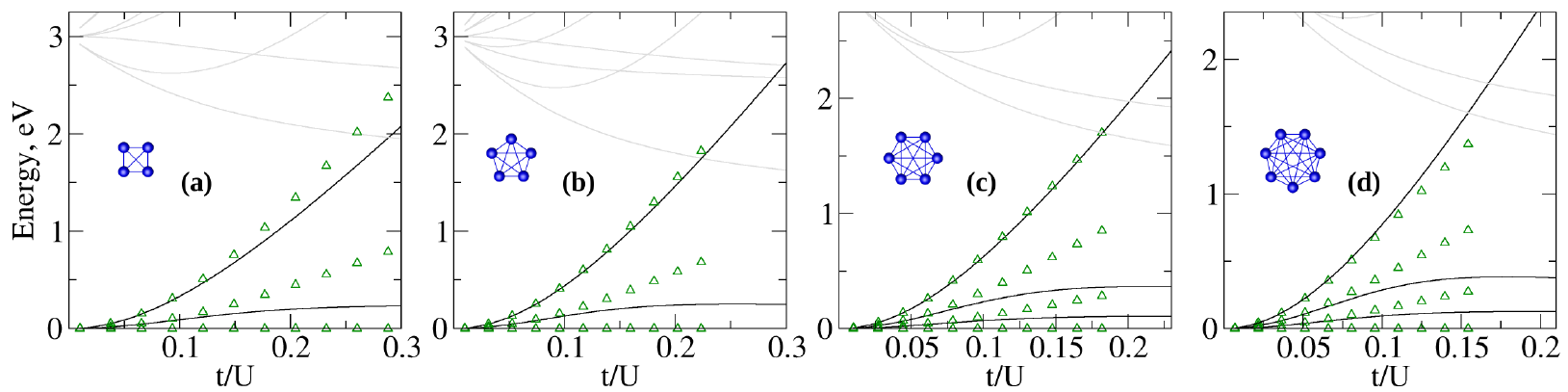}
\caption {Excitation spectra of all-to-all systems. Solid lines denote the Hubbard model spectrum. Green triangles correspond to the Heisenberg model solutions with the exchange interactions calculated by the high-temperature expansion method, Eq.\ref{eq:J}.}\label{fig:pyramids}
\end{figure*}

The simplest Heisenberg Hamiltonian with two-spin exchange interactions constructed by the high-temperature expansion method gives the eigenvalue spectrum that is coincident with the Hubbard model one up to $t/U$ = 0.07 (Fig.\ref{fig:pyramids}). For larger values of $t/U$ we observe the deviation of the spin and electronic models that mainly concerns intermediate excited levels. The problem may be resolved by introducing the high-order multispin interactions (four-spin and six-spin).\cite{Mila}

\begin{figure}
\includegraphics[angle=0,scale=1]{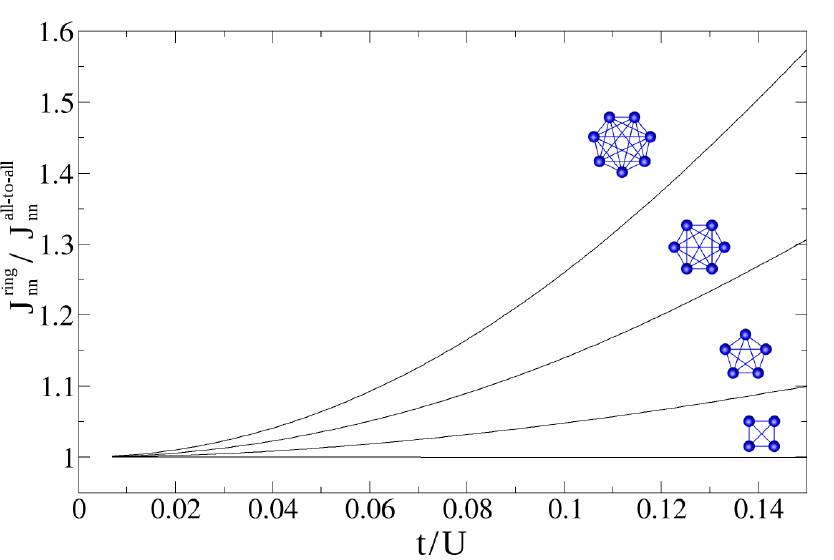}
\caption {The calculated ratio $\frac{J^{ring}_{nn}}{J^{all-to-all}_{nn}}$ demostrating the contribution of the indirect hopping processes to the two-spin interaction.}\label{fig:pyrs_draft}
\end{figure}

The pair exchange interaction between nearest neighbors in a quantum ring has the direct contributions, proportional to $t_{ij}t_{ji}$ and high-order non-direct ones, $\sum_{k} t_{ik}t_{kj}$, where the site index $k \ne i,j$.  
In case of the configurations with all-to-all hoppings the non-direct processes become very efficient and strongly contribute to the exchange interaction between two spins. It can be seen from Fig.\ref{fig:pyrs_draft}. For each pair in the N-site ring there are N-2 non-direct exchange path including one intermediate site. 

\section{Conclusion}
We propose the method for calculations of the magnetic interactions in the paramagnetic phase. Being formulated in the high-temperature and localized spin limits our approach can be used for constructing the spin Hamiltonian in a wide range of the $t/U$ ratios. It was shown by the classical examples such as the dimer and triangle finite clusters.
By using the proposed method we investigated the magnetic couplings in quantum spin rings with different hopping configurations. 
Our methodological and calculation results will be useful for analysis of the data obtained in experiments with ultracold fermions that provide unique possibility to measure and control the spin-spin correlation function between two sites in optical lattice.\cite{Greif1}
The proposed scheme can be also applied for simulating the magentic couplings between impurities in metallic host. For that instead of the Hubbard model one should solve two-impurity Anderson model.

\section{Acknowledgments.}
We acknowledge fruitful communications with Mikhail Katsnelson, Alexander Lichtenstein, Andrea Secchi, Alexander Tsirlin, Andrey Katanin, Sergey Brener, Igor Solovyev and Vladimir Anisimov.
The work is supported by the grant program of the Russian Science Foundation 15-12-20021.

\end{document}